\documentstyle[epsfig,epsf,ifthen,multicol,aps,prb]{revtex}

\begin{document}



\title{Electron-electron interactions in a weakly screened two-dimensional electron system}

\author{I. Karakurt and A.J. Dahm}
\address{Department of Physics, Case Western Reserve University, Cleveland, \\OH 44106-7079,USA}

\setlength{\voffset}{1cm}
\date{\today}

\maketitle

\begin{abstract}
We probe the strength of electron-electron interactions using magnetoconductivity measurements of two-dimensional 
non-degenerate electrons on liquid helium at 1.22 K. Our data extend to electron densities that are two orders of 
magnitude smaller than previously reported.  We span both the independent-electron regime where the data are 
qualitatively described by the self-consistent Born approximation (SCBA), and the strongly-interacting electron 
regime.  At finite fields we observe a crossover from SCBA to Drude theory as a function of electron density.


\end{abstract}


\begin{multicols}{2}

Electrons supported by a liquid-helium surface form a low density, non-degenerate two-dimensional (2D) electron gas.  
Aside from the non-degeneracy, it differs from other 2D electron systems in the strength of the electron-electron (e-e) 
interaction\cite{Dykman0,Lea2,Lea3,Dykman1,Lea1}.  The Coulomb interaction is weakly screened by metallic plates that 
are separated from the electron layer by about 1 mm.  It is an ideal system for testing the properties of 
strongly-interacting electrons.  

One of the interesting properties of this non-degenerate 2D electron gas is the density of states (DOS) in a magnetic 
field.  The DOS peaks at the Landau levels (LLs) have a width that depends both on the scattering rate and the e-e 
interaction.  The width of the DOS peaks has been calculated\cite{Ando1,Ando2} in the self-consistent Born 
approximation (SCBA) and has been studied experimentally\cite{Lea2,Lea3,Dykman1,Lea1,Heijden,Neuenschwander,Scheuzger2,Scheuzger} 
through measurements of the magnetoconductivity.  We report magnetoconductivity measurements from an extremely low 
density, $\sim 1.9$ x$10^9$ m$^{-2}$, where e-e interactions are negligible, to densities where Coulomb interactions 
dominate the width of the DOS peaks.  

In this system, electron-helium atom scattering dominates at temperatures above $0.8$ K, while electron-ripplon 
scattering is important at lower temperatures.  Each collision between an electron and a helium atom changes the 
electron energy by about one percent.  Thus, scattering is quasi-elastic.  The helium atoms in the vapor act as short 
range-scatterers to a very good approximation.

In zero magnetic field the density of states is constant: $D_0(E)=m/\pi\hbar^2$.  The magnetoconductivity of electrons, 
$\sigma_{xx}(B)$, is given in the Drude model for $\mu_0B \ll 1$ as
\begin{equation}
\sigma_{xx}(B)=\frac{\sigma_{xx}(0)}{[1+(\mu_0B)^2]},\; \; \; \; \;  \sigma_{xx}(0)=ne\mu_0,
\end{equation}
where $\sigma_{xx}(0)$, $\mu_0=e\tau_0/m$, and $\tau_0$ are the zero field conductivity, mobility and scattering time, 
respectively.  The Drude model assumes that electrons are independent and move in straight paths (in zero field) and in 
classical orbits (in a magnetic field).  

A magnetic field transverse to the 2D electrons changes the density of states, and consequently the magnetoconductivity 
of electrons dramatically.  In a magnetic field, Landau levels separate when $\Delta/\hbar \omega_c \sim 1$.  Here, 
$\Delta$ is the width of the Landau level.  The broadening $\Delta_a$ due to collisions with helium atoms and the 
broadening $\Delta_e$ due to electron-electron interactions enter the $\Delta$ as\cite{Lea1}
\begin{equation}
\Delta^2= \Delta_a^2+\Delta_e^2.
\end{equation}
Once LLs form, the Drude model loses its validity and a crossover to the SCBA regime occurs.  In SCBA, the DOS and thus 
the magnetoconductivity is obtained self consistently in the Born approximation.  The broadening $\Delta_a$ has been 
calculated\cite{Ando1,Ando2} in the SCBA limit for a semi-elliptic DOS and short range scatterers and given by
\begin{equation}
\Delta_a=\frac{\hbar}{\tau_B}= \hbar(\frac{8}{\pi}\frac{\omega_c}{\tau_0})^{1/2},
\end{equation}
where $\omega_c=eB/m$ is the cyclotron frequency and $\tau_B^{-1}$ is the scattering rate in a magnetic field. For 
$\Delta_e \rightarrow 0$ and $\hbar \omega_c < \Delta_a$, we assume that the broadening $\Delta_a$ is determined by 
the zero field scattering time and is on the order of $\sim \hbar/\tau_0$.  The crossover is delayed by many electron 
effects\cite{Lea2,Dykman1,Lea1} as seen in Eq. 2.  The broadening $\Delta_e$ is given by theory\cite{Lea1} as
\begin{equation}
\Delta_e= eE_f\lambda_T;\; \; \; \; \;
E_f \approx (\frac{11kTn^{3/2}}{4 \pi \overline{\epsilon} \epsilon_0 })^{1/2},
\end{equation}
where $\overline{\epsilon}=(\epsilon_{He}+1)/2=1.028$, $E_f$ is the fluctuating field\cite{Dykman0,Lea2} an electron 
feels due to redistribution of other electrons as it moves, and the thermal wavelength $\lambda_T$ is the characteristic 
size of an electron in the classical limit $\hbar \omega_c < kT$, which holds for our experimental data.

The SCBA is valid for static scatterers and a vanishing coherence time\cite{Ando2,Kuehnel}.  For electrons on helium 
the coherence time is on the order of $\tau_0$, and the scattering is dynamic, i.e., the vapor atoms move slowly.  For 
this case the behavior should be qualitatively given by the SCBA expression.

The crossover to the SCBA regime has already been confirmed by earlier experiments in the quantum limit where 
$\hbar \omega_c > kT$, whereas no such crossover has been reported in the classical limit.  This resulted from the 
electron densities being too high (greater than $10^{11}$ m$^{-2}$) in previous 
experiments.\cite{Lea2,Dykman1,Heijden,Neuenschwander,Scheuzger2,Scheuzger}  At those densities the LLs  can separate 
only at very high magnetic fields due to many electron effects.  In our experiment, we were able to reduce the electron 
density another two orders of magnitude to $1.9\times10^{9}$ m$^{-2}$.  At this density, the plasma parameter 
$\Gamma =e^2(\pi n)^{1/2}/4\pi \epsilon_0kT$, which is the ratio of the unscreened Coulomb energy to the thermal 
kinetic energy, is $\sim1.1$ compared to $\sim13$ at $n \sim 10^{11}$ m$^{-2}$.  Thus we were able to observe, for 
the first time, the SCBA magnetoconductivity in classical fields $\hbar \omega_c < kT$.

In Fig 1. we give a qualitative picture of broadening of a semi-elliptic LL by electron-electron interactions.  
The quantity $\Delta_e$ is, in fact, the uncertainty in the energy of an electron due to its finite size $\lambda_T$ 
in a field $E_f$ and depends strongly on the electron density via $E_f$.  The theory predicts that the broadening 
$\Delta_e$ is on the order of $\Delta_a$ for $\mu_0=25$ m$^2$/Vs and $n \sim$ 10$^{11}$ m$^{-2}$.
\begin{figure}
\begin{center}
\includegraphics[width=1\linewidth]{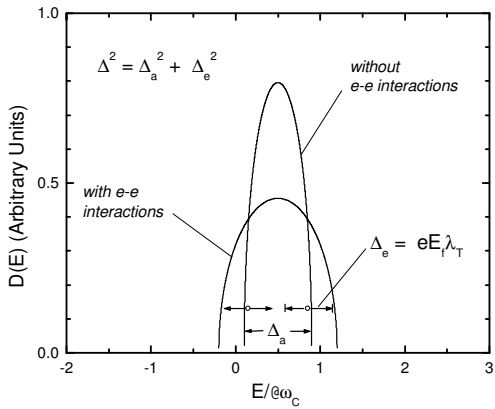}
\caption{A qualitative picture of broadening of a LL due to many electron effects.  The additional uncertainty 
$\Delta_e$, depicted by double arrows in the figure, broadens the width. The total broadening is given by $\Delta$.}
\label{fig1}
\end{center}
\end{figure}
An expression for the magnetoconductivity of 2D non-degenerate electrons has been calculated by van der Heijden 
et. al.\cite{Heijden} in SCBA regime and given for a semi-elliptic density of states by
\begin{eqnarray}
\nonumber \sigma_{xx}(B)= \frac{2 \coth(\hbar \omega_c/kT)}{\pi I_1(\Delta_a/2kT)}f(\Delta_a,T);\\
f(\Delta_a,T)=[\cosh(\frac{\Delta_a}{2kT})-\frac{\Delta_a}{2kT}\sinh(\frac{\Delta_a}{2kT}) ]\frac{ne}{B},
\end{eqnarray}
where $I_1$ is the modified Bessel function of order 1, $n$ is the electron density, and $\Delta_a$ is the width of 
Landau levels.  Eq. 5 is valid when $\Delta_e \ll \Delta_a$, i.e. when $\Delta \rightarrow \Delta_a$.

For $\hbar \omega_c, \Delta_a \ll kT$, Eq. 5 gives
\begin{equation}
\frac{\sigma_{xx}(0)}{\sigma_{xx}(B)}= \frac{3\pi^{3/2}}{8 \sqrt{2}} \; (\mu_0 B)^{3/2}.
\end{equation}
Here, we emphasize the $B^{3/2}$ dependence of the SCBA magnetoresistivity compared to the $B^2$ dependence of the 
Drude magnetoresistivity.

We measured the inverse magnetoconductivity $1/\sigma_{xx}(B)$ of electrons as a function of a magnetic field 
perpendicular to the 2D electron layer in a Corbino geometry.  Electrons were deposited over a $\sim1$ mm thick 
helium film from a gaseous discharge.  The density of electrons was controlled by carefully adjusting the dc holding 
voltage on the Corbino electrodes below the liquid helium surface. 
\begin{figure}
\begin{center}
\includegraphics[width=1\linewidth]{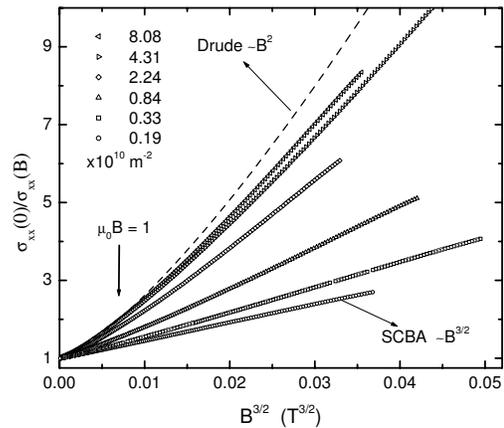}
\caption{Normalized inverse-magnetoconductivity plotted as a function of $B^{3/2}$ at $T=1.22$ K.  The densities are 
in units of $10^{10}$ m$^{-2}$, and $\mu_0=27.5$ m$^2$/Vs.}
\label{fig2}
\end{center}
\end{figure}
In Fig 2., we show the normalized inverse magnetoconductivity $\sigma_{xx}(0)/\sigma_{xx}(B)$ as a function of 
$B^{3/2}$ for six electron densities.  We see a crossover from the SCBA magnetoconductivity ($B^{3/2}$ dependence) 
to the Drude magnetoconductivity ($B^2$ dependence) as the electron density is increased for $\hbar \omega_c/\Delta > 1$.  
The dashed line is the normalized theoretical Drude magnetoconductivity calculated using the experimental parameters 
of the highest density curve.  We obtain the zero-field mobility and the density for the highest density curve from 
a fit to the Drude theory in small fields.  Then, we calculate the electron densities for the lower density curves 
using the measured zero-field resistivitiy $1/ne\mu_0$ and assuming the same zero-field mobility of $\mu_0=27.5$ m$^2$/Vs 
obtained for the highest density curve.  The fact, that the magnetic field region $\hbar \omega_c/\Delta < 1$ within 
which the Drude-like behavior is observed gets smaller as the electron density is reduced, prevented us from obtaining 
the zero-field mobilities accurately and the densities independently for the lower density curves.  

The effect of LL broadening due to e-e interactions is seen clearly in the figure as the crossover to the SCBA regime 
occurs at a lower magnetic field as the electron density is reduced.  The $B^{3/2}$ dependence is clear in 
the curve with the lowest electron density of $0.19\times10^{10}$ m$^{-2}$.  Note that, in our system, the quantum 
limit is reached when $B = 0.91$ Tesla.  The values of $\hbar \omega_c / kT$ for all the data shown were less than 
0.12, and the magnetic fields used fall well into the classical regime.
\begin{figure}
\begin{center}
\includegraphics[width=1\linewidth]{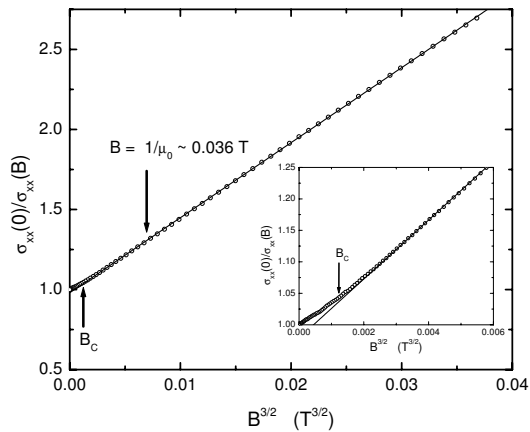}
\caption{Normalized inverse-magnetoconductivity vs. $B^{3/2}$ for the lowest density $n=0.19\times10^{10}$ m$^{-2}$.  
$T=1.22$ K. }
\label{fig3}
\end{center}
\end{figure}
We plot the curve with the lowest density in Fig. 3.  Except for very low fields the data follow SCBA.  The field 
$B$ at which $\mu_0 B =1$ and the crossover field $B_c$ are shown with arrows. We explain how we obtain the field 
$B_c \sim 0.011$ T later in the text.  The crossover is seen more clearly in the inset of the figure where we show 
a blow up of the same graph at small fields.  The field $B_c$ corresponds to $\sim 30\%$ of the theoretical value 
$1/\mu_0 \sim 0.036$ T given for no e-e broadening.  The solid line is the theoretical fit to Eq. 6.  The fit 
gives a zero field mobility of $\sim$10 m$^2$/Vs which is smaller than the value $27.5$ m$^2$/Vs.  This is in 
agreement with a lower value of the mobility obtained from fits to the SCBA formula in earlier measurements 
reported in the quantum limit.\cite{Heijden}

Our data would appear to disagree with theory, which predicts that the contribution $\Delta_e$ to the total broadening 
$\Delta$ becomes negligible at electron densities below $10^{11}$ m$^{-2}$ and that the crossover to SCBA should occur 
at fields $\sim 1/\mu_0$.  The SCBA magnetoconductivity is not fully developed at $B=0.1$ T even at the low density of 
$0.84\times10^{10}$ m$^{-2}$ as a $\ln(\sigma_0/\sigma_{xx})$ vs. $\ln(\mu_0B)$ graph yields a slope of $\gtrsim 1.7$ which is 
still greater than the value of $1.5$ in SCBA. The contribution $\Delta_e$ to the total 
broadening is more significant than predicted by theory.  

In order to obtain a quantitative result for the crossover field, we plot our data as $[\sigma_{xx}(0)/\sigma_{xx}(B)]-1$ 
and fit it with the following function $F(\mu_0,B,B_c)$
\begin{eqnarray}
\nonumber F(\mu_0,B,B_c) = [1-C(B,B_c)](\mu_0 B)^2\\
+\frac{3\pi^{3/2}}{8 \sqrt{2}} \; C(B,B_c)(\mu_0 B)^{3/2},
\end{eqnarray}
where $\mu_0$ and $B_c$ are the free parameters. The $C(B,B_c)$ is a rapidly-changing crossover function and we find an 
excellent fit to the low-density data by choosing
\begin{equation}
C(B,B_c) = \tanh^{1/2}(\frac{B}{4B_c}).
\end{equation}  
This function is $0.5$ at $B=B_c$. The fitting function $F$ starts in the Drude regime at $B=0$ and goes into SCBA at a 
magnetic field characterized by $B_c$:
\begin{eqnarray}
\nonumber F(\mu_0,B,B_c) = (\mu_0 B)^2 \;\;\;, B \ll B_c\\
=\frac{3\pi^{3/2}}{8 \sqrt{2}} \;(\mu_0 B)^{3/2}\;\;\;, B \gg B_c.
\end{eqnarray}
The values of $B_c$ obtained from the fits are a measure of the actual crossover field and thus the width $\Delta$, 
but do not necessarily represent the actual widths of the LLs, since the overlap between neighboring levels at the actual 
crossover field is not known exactly and the fitting function has no theoretical basis.
\begin{figure}
\begin{center}
\includegraphics[width=1\linewidth]{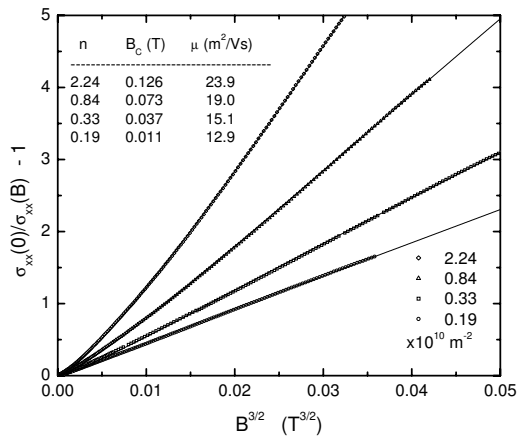}
\caption{The four lowest-density-curves with fits to the function $F(\mu_0,B,B_c)$ vs. $B^{3/2}$.}
\label{fig4}
\end{center}
\end{figure}
In Fig. 4., we show the curves with their non-linear least square fits, shown by the solid lines, to Eq. 7.  The best 
fits are obtained allowing $\mu_0$ and $B_c$ to be free parameters.  We leave the curves with the highest two 
densities out of this analysis since, for those data, the crossover region extends too far beyond the magnetic fields 
used in the experiments, and this results in an error in the fitting parameter $B_c$.  

The values of $B_c$ obtained from the fits above give an approximate width $\Delta^\star$ of the LLs for each electron 
density.  We set $\Delta^\star=\hbar \omega_c(B=B_c)=\hbar e B_c/m$, and plot the values of $\Delta^\star$ as a function of 
electron density in Fig. 5.  In order to compare with the theoretical expression for the width of the LLs, we 
plot Eq. 2 as a solid line in the figure for $\Delta_a=15$ mK and $\Delta_e=11  eE_f \lambda_T$.  Although the 
values of $\Delta^\star$ obtained from the fits differ from the theoretical values by a factor of $11$, they give 
the correct functional dependence on the electron density. 
\begin{figure}
\begin{center}
\includegraphics[width=1\linewidth]{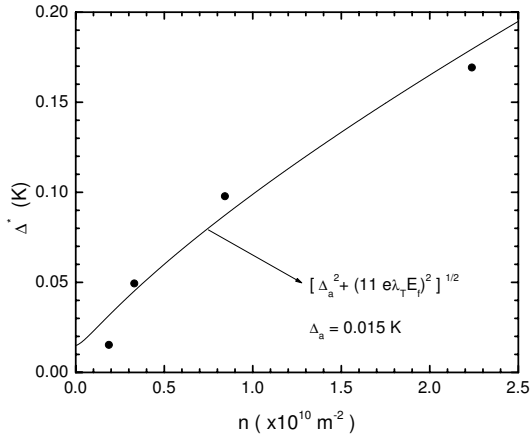}
\caption{The values of $\Delta^\star$ as a function of electron density.  The solid line is described in the text.}
\label{fig5}
\end{center}
\end{figure}
The values of $\mu_0$ from the fits correspond to a reduced
mobility for low densities.  In our initial fits to the data we allowed for different mobilities in the two regimes 
as adjustable parameters. The best fits were given with equal mobilities for the
two regimes. At higher densities van der Heijden et
al.\cite{Heijden} deduced a mobility from the SCBA regime that was
a factor of two smaller than the Drude mobility.  However, our
fits required an adjustment of the Drude mobility as well.  The
smaller apparent mobility occurs in part, because the low-density
data is weighted by the SCBA regime, and we observe no
well-developed Drude regime at low densities.  Deviations from
Drude behavior begin at zero field in our data. Despite this fact,
the strong variation in the field where the data crossover to the
SCBA regime is apparent in both the data curve shown in Fig. 2 and
the fits shown in Fig. 4.  The crossover function $C(B,B_c)$ under-weighs the Drude part, but fits very well because 
there is no well developed Drude regime at very low densities in our data.

In conclusion, we measured the magnetoconductivity of
non-degenerate electrons in the very low-density limit.  The
effect of e-e interactions is clearly demonstrated in these data.
Electron-electron interactions have a significant effect on the
magnetoconductivity causing a delay in the transition from Drude
to SCBA regime as a function of a magnetic field.  When the many
electron effects are negligible the transition is observed in
classical fields.  We studied the dependence of the LL width on
the electron density for the first time.  Our results agree with
the theory qualitatively, but differ by a large numerical factor.  A more detailed theoretical analysis of 
classical electron scattering from dynamic scatterers in a magnetic field is required to reconcile experiment and theory.

\section*{ACKNOWLEDGMENTS}
The authors wish to acknowledge M.I. Dykman, D. Herman, and H. Mathur for helpful conversations.  This work was 
supported in part by NSF grant DMR-0071622.

\noindent
\vspace{.5 cm}

\bibliographystyle{prsty}

\begin{references}

\bibitem{Dykman0} M.I. Dykman and L.S. Khazan, {\it Sov. Phys. JETP} {\bf 50(4)}, 747 (1979).

\bibitem{Lea2} M.J. Lea and M.I. Dykman, {\it Philos. Mag. B} {\bf 69}, 1059 (1994).

\bibitem{Lea3} M.J. Lea, P.Fozooni, P.J. Richardson, and A. Blackburn,  {\it Phys. Rev. Lett. } {\bf 73}, 1142 (1994).

\bibitem{Dykman1} M.I. Dykman, M.J. Lea, P.Fozooni, and J. Frost, {\it Physica B} {\bf 197}, 340 (1994).

\bibitem{Lea1} M.J. Lea, P.Fozooni, A. Kristensen, P.J. Richardson, K.
      Djerfi, M.I. Dykman, C.Fang-Yen, and A. Blackburn, {\it Phys. Rev. B } {\bf 55}, 16280 (1997).

\bibitem{Ando1} T. Ando, {\it J. Phys. Soc. Jpn. } {\bf 37}, 1233 (1974).

\bibitem{Ando2} T. Ando and Y. Uemura, {\it J. Phys. Soc. Jpn. } {\bf 36}, 959 (1974).

\bibitem{Heijden} R.W. van der Heijden, M.C.M. van de Sanden, J.H.G. 
     Surewaard, A.T.A.M. de 	 Waele, H.M. Gijsman, and F.M. 
     Peeters, {\it Europhys. Lett.} {\bf 6}, 75 (1988).

\bibitem{Neuenschwander} J. Neuenschwander, P. Scheuzger, W. Joss, and P. Wyder, {\it Physica  B } {\bf 165-166}, 845 (1990).

\bibitem{Scheuzger2} P. Scheuzger,J. Neuenschwander, and P. Wyder, {\it Helv. Phys. Acta. } {\bf 64}, 170 (1991).

\bibitem{Scheuzger} P. Scheuzger, J. Neuenschwander, W. Joss, and P. Wyder, {\it Physica  B } {\bf 194-196}, 1231 (1994).

\bibitem{Kuehnel} Frank Kuehnel, Leonid P. Pryadko, and M.I. Dykman, Cond-Mat/0001427.


\end{references}

\end{multicols}

\end{document}